\documentclass[manuscript,screen]{acmart}

\AtBeginDocument{%
  \providecommand\BibTeX{{%
    \normalfont B\kern-0.5em{\scshape i\kern-0.25em b}\kern-0.8em\TeX}}}

\copyrightyear{2024} 
\acmYear{2024} 
\setcopyright{rightsretained} 

\begin{document}

\title{Large Language Models and Video Games: A Preliminary Scoping Review}

\author{Penny Sweetser}
\email{penny.kyburz@anu.edu.au}
\affiliation{%
  \institution{The Australian National University}
  \city{Canberra}
  \country{Australia}
}

\renewcommand{\shortauthors}{Penny Sweetser}

\begin{abstract}
Large language models (LLMs) hold interesting potential for the design, development, and research of video games. Building on the decades of prior research on generative AI in games, many researchers have sped to investigate the power and potential of LLMs for games. Given the recent spike in LLM-related research in games, there is already a wealth of relevant research to survey. In order to capture a snapshot of the state of LLM research in games, and to help lay the foundation for future work, we carried out an initial scoping review of relevant papers published so far. In this paper, we review 76 papers published between 2022 to early 2024 on LLMs and video games, with key focus areas in game AI, game development, narrative, and game research and reviews. Our paper provides an early state of the field and lays the groundwork for future research and reviews on this topic.

\end{abstract}

\ccsdesc[500]{Software and its engineering~Interactive games}
\ccsdesc[500]{Computing methodologies~Natural language generation}
\ccsdesc[500]{Applied computing~Computer games}

\keywords{large language models, LLMs, games, videogames, GPT}

\maketitle

\section{Introduction}

Since OpenAI's release of ChatGPT in late 2022, awareness and usage of Large Language Models (LLMs) has surged in research, development, and the general population across a broad spectrum of domains. LLMs are powerful tools for language processing and prediction, pre-trained on vast collections of natural language, and capable of performing diverse language analysis and generation tasks \cite{Hadi_2023}. The release of ChatGPT, along with the many other available LLMs (e.g., GPT-4, LLaMa, Codex, BERT) has opened new doors to research and development potential, which has seen a recent increase in related research. Like many fields, LLMs hold interesting possibilities for video games, which has prompted many researchers to hasten to investigate the potential for applying LLMs to various aspects of video game research and development. Although the concept of generative AI is not new to video games, with decades of prior work in AI-powered generation of game content \cite{shaker2016procedural, hendrikx2013procedural}, LLMs have the potential to revolutionise generation and co-creation of video game content, along with game development tools and processes, and games research approaches. As research and development of LLMs and games is occurring and evolving quickly, it is difficult to capture a full picture of how LLMs are being used in games research. The aim of this paper is to provide a preliminary scoping review of LLMs and video games, surveying the related research conducted between 2020 and 2023. We aim to identify the ways in which researchers have been exploring the use of LLMs for game development and research to date. To identify the relevant papers, we conducted a Google Scholar search for papers published between 2020-2023 (and very early 2024). We identified 76 relevant papers from 2260 results returned in the search. Our review of the papers revealed that 27/76 (35.5\%) of the papers related to game AI, 25/76 (32.9\%) to game development, 17/76 (22.4\%) to narrative, and 7/76 (9.2\%) to game research and reviews. We additionally found another 9 papers that used game data sets as part of broader recommendation research. In this paper, we analyse the work conducted using LLMs for games research to date, providing a snapshot of the state of the field and a grounding for future research in this area.

\begin{table*}
  \caption{Relevant papers on video games and LLMs by theme.}
  \begin{tabular}{lp{10.5cm}}
    \toprule
    \textbf{Theme} & \textbf{Papers}\\
    \midrule
    Game AI and Agents  &  \cite{croissant2023appraisal, lin2023mm, ma2023large, li2023assessing, kaiya2023lyfe, lawley2023val, colas2023augmenting, piterbarg2023diff, park2023generative, zhu2023ghost, wang2023voyager, du2023guiding, nottingham2023selective, nottingham2023selective2, klissarov2023motif, leecan, wu2023spring, yang2023octopus, yan2023ask, xu2023language, agashe2023evaluating, guan2023efficient, chen2023agentverse, brawer2023towards, verma2023preference, liu2023llm, gong2023mindagent} \\
    \midrule
    Game Development and Play  &  \cite{ren2023make, todd2023level, sudhakaran2023prompt, nasir2023practical, kumaran2023end, awiszus2022wor, roberts2022steps, roberts2022surreal, sudhakaran2023mariogpt, taveekitworachai2023chatgpt4pcg, chen2023gamegpt, shams2023towards, colado2023using, saito2023double, tinterri2024ai, junior2023chatgpt, bottega2023jubileo, yue2023combine, amresh2023integrating, lanzi2023chatgpt, huang2023create, zhu2023fireball, gongora2023skill, taesiri2023glitchbench, taesiri2022large} \\
    \midrule
    Narrative, Story, and Dialogue  &  \cite{nimpattanavong2023fighting, csepregi2021effect, latouche2023generating, muller2023chatter, akoury2023towards, akoury2023framework, volum2022craft, sun2023fictional, paduraruconversational, kumaran2023scenecraft, ashby2023personalized, taveekitworachai2023waiting, gursesli2023chronicles, taveekitworachai2023journey, yong2023playing, sun2023language, vartinen2022generating} \\
    \midrule
    Game Research and Reviews  &  \cite{lankes2023game, viggiato2023leveraging, de2023performing, svetasheva2024harnessing, hamalainen2023evaluating, hamalainen2022neural, ramirez2023controlling} \\
    \midrule
    Recommendation  &  \cite{yue2023llamarec, yang2023large, zhang2023recommendation, bao2023bi, lei2023recexplainer, zheng2023adapting, shu2023rah, sun2023large, charity2023preliminary} \\
  
  \bottomrule
\end{tabular}
\label{tab:papers}
\end{table*}

\begin{figure*}
  \centering
  \includegraphics[width=1.0\linewidth]{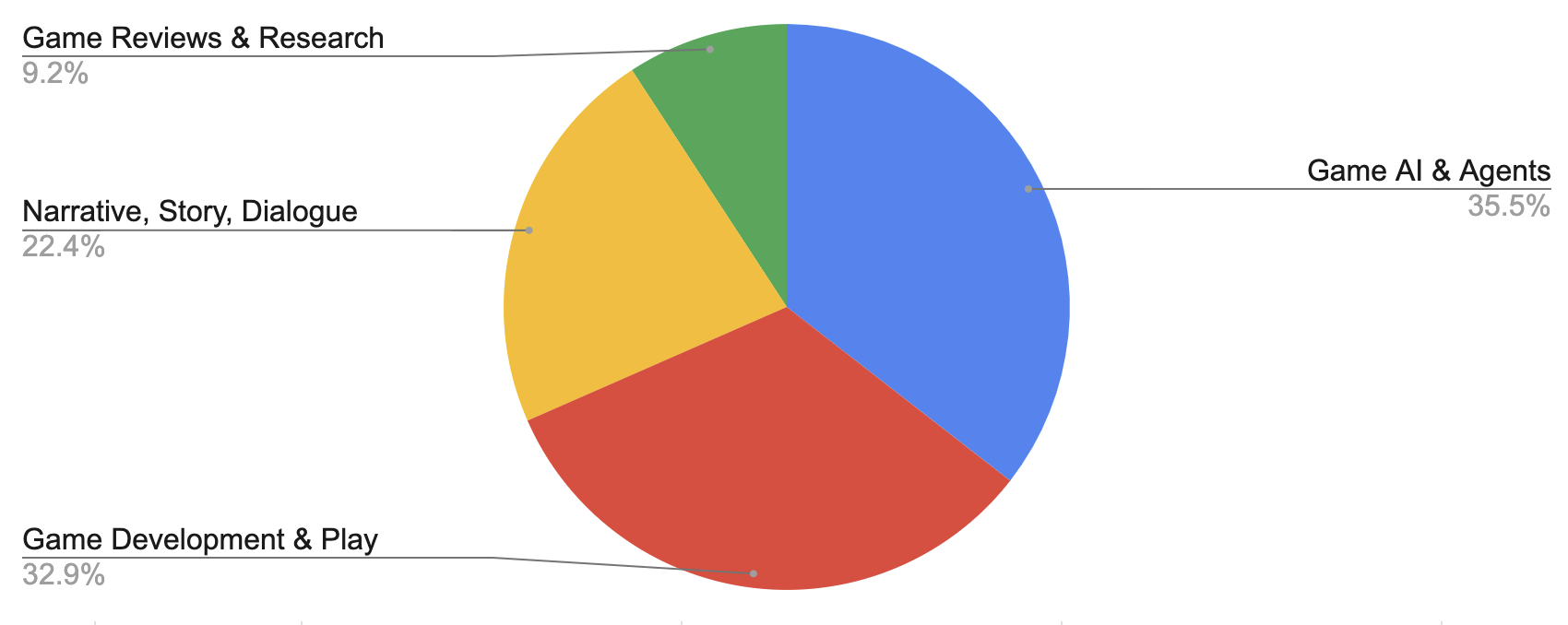}
  \caption{Percentage of relevant papers identified in search by theme.}
  \Description{.}
  \label{fig:chart}
\end{figure*}

\section{Scoping Review}

In order to carry out our scoping review, we conducted a Google Scholar search for papers related to LLMs and video games published between 2020 to 2023 (and very early 2024). We used Google Scholar as it includes a wide variety of sources and publication types and allows full text search. We conducted our search on 15 January 2024, using the search string "(LLM OR LLMs OR "large language model" OR "large language models") AND ("video game" OR "video games" OR videogame OR videogames)" for the period 2020 to 2024. We used a full text search (rather than title or abstract only). Our search returned 2260 results, which we sorted by relevance. We reviewed the first 500 returned records to identify relevant papers. We found that the majority (80\%) of relevant papers were in the first 250 records, with relevance rapidly decreasing after this point. 

To identify relevant papers, we reviewed the title and abstract of each paper, and searched the full text for "game" and "large language model" to find the usage. Most papers mentioned games in previous work or as an example, but did not include games in their own research. Some papers used the acronym "LLM" for another meaning. We considered papers relevant if they reported on original research related to video games and LLMs, either with a focus on games or using games as a testbed. We excluded general reviews that mentioned games and broader work that could be applied to games but that had not been applied to games in the paper. Through our search, we identified 71 relevant papers from the 2260 results. We found an additional five relevant papers while reviewing the 71 identified papers, for a full set of 76 papers. The majority (68/76, 90.7\%) of papers were published (including preprints) in 2023, with 2 (2.7\%) in 2024 and 6 (8.0\%) in 2022. This suggests that any future searches with the same focus could start from 2022. A search with the same search string from 2022 onwards returned 1860 records (400 fewer). 

During our search, we coded our identified relevant papers into broad categories, so that we could further review the similar papers together and identify the key themes. Following the initial coding, we finalised 4 key themes that captured all papers and recoded all papers into these themes. The most common theme was "Game AI and Agents" with 27/76 (35.5\%) of the papers. The second theme was "Game Development and Play" with 25/76 (32.9\%) the papers. The third theme was "Narrative, Story, and Dialogue" with 17/76 (22.4\%). The fourth and final theme, "Game Research and Reviews", captured the remaining 7/76 (9.2\%) of the papers. We additionally found another 9 papers that used game data sets as part of broader recommendation research. Although we did not consider this to fit the focus of our search, we believe this related topic is of interest and relevant to future work on games and LLMs, so we also captured these papers for analysis.

The final step in our scoping review was to read and analyse each of the relevant papers. The following sections present our analysis of the papers, grouped by each of our key themes. We summarise the focus of the research conducted under each of these themes, extract sub-themes, and discuss relevant findings. Across our key themes, we found that by far the most commonly used LLM was GPT (65/76, 85.5\%), followed by LLaMA (7/76, 9.2\%), and Codex and BERT (each 5/76, 6.6\%), with some papers using multiple LLMs.

\subsection{Game AI and Agents}

Over a third (27/76, 35.5\%) of the relevant papers identified in our search related to the theme of Game AI and Agents. The papers within this theme related to game agent design and behaviours (11/27) \cite{croissant2023appraisal, lin2023mm, ma2023large, li2023assessing, kaiya2023lyfe, lawley2023val, colas2023augmenting, piterbarg2023diff, park2023generative, zhu2023ghost, wang2023voyager}, LLMs and reinforcement learning (9/27) \cite{du2023guiding, nottingham2023selective, nottingham2023selective2, klissarov2023motif, leecan, wu2023spring, yang2023octopus, yan2023ask, xu2023language}, and collaboration and coordination (7/27) \cite{agashe2023evaluating, guan2023efficient, chen2023agentverse, brawer2023towards, verma2023preference, liu2023llm, gong2023mindagent}.

The papers on game agent design and behaviours experimented with applying LLMs to different aspects of agents, including reasoning \cite{li2023assessing, kaiya2023lyfe}, playing games \cite{lin2023mm, ma2023large}, embodied agents \cite{piterbarg2023diff, wang2023voyager}, goals and tasks \cite{colas2023augmenting, lawley2023val}, emotion simulation \cite{croissant2023appraisal}, believable agents \cite{park2023generative}, and general AI \cite{zhu2023ghost}. The papers on reasoning tested LLM capabilities at logical reasoning via Minesweeper \cite{li2023assessing} and social reasoning in a custom multi-agent environment \cite{kaiya2023lyfe}. \citet{li2023assessing} found that LLMs possessed foundational abilities, but that they could not integrate these into a multi-step logical reasoning process to solve Minesweeper. \citet{kaiya2023lyfe} argued that their Lyfe Agents exhibited human-like self-motivated social reasoning, which could enrich human social experiences in virtual worlds. The papers on playing games applied GPT-4V(ision) to analyse video to predict next steps when playing a Mario video game \cite{lin2023mm} and to test the capabilities of LLMs to play StarCraft II (SC2) \cite{ma2023large}. \citet{lin2023mm} claimed that their agent displayed an understanding of the Mario game and generated reasonable action controls to play the game effectively. \citet{ma2023large} concluded that LLMs possess the relevant knowledge and complex planning abilities to play SC2 (in their textual SC2 environment). The papers on embodied agents investigated the use of diff history for environment observations in NetHack \cite{piterbarg2023diff} and an LLM-powered embodied lifelong learning agent (Voyager) in Minecraft \cite{wang2023voyager}. \citet {piterbarg2023diff} found that using diff history allowed observational compression and abstraction that yielded a 7x improvement in game score and outperformed agents that use visual observations by over 40\%. \citet{wang2023voyager} argued their their Voyager agent shows strong in-context lifelong learning capability and exceptional proficiency at playing Minecraft. The paper on tasks \cite{lawley2023val} presented an interactive task learning system (VAL) that uses LLMs only for specific tasks to support interactive learning of hierarchical task knowledge from natural language. They found that the acquired knowledge was human interpretable and generalised to support execution of novel tasks without additional training and that users could successfully teach VAL using natural language in the Overcooked-AI video game setting. The paper on goals \cite{colas2023augmenting} introduced an LLM-based agent to support the representation, generation and learning of diverse, abstract, human-relevant goals in a text-based environment, Cooking World. They found that their agent learned to master a large diversity of skills without reward functions, curriculum, or hand-coded goal representations. The remaining papers investigated using LLMs for agents to solve emotional intelligence tasks and to simulate emotions in a conversational RPG game \cite{croissant2023appraisal}, to synthesise memories over time into high-lever reflections used to plan behaviour in a Sims-like environment \cite{park2023generative}, and to create generally capable agents in Minecraft \cite{zhu2023ghost}. \citet{croissant2023appraisal} concluded that their work provided a first step towards better affective agents represented in LLMs. \citet{park2023generative} argued that their generative agents produced believable individual and emergent social behaviors. Finally, \citet{zhu2023ghost} concluded that their agent demonstrated the potential of LLMs in developing capable agents for handling long-horizon, complex tasks, and adapting to uncertainties in open-world environments.

The papers on reinforcement learning (RL) investigated combining LLMs with RL in game environments, focusing on state descriptions \cite{nottingham2023selective, nottingham2023selective2}, exploration and intrinsic motivation \cite{du2023guiding, klissarov2023motif}, chain-of-thought reasoning \cite{wu2023spring, yan2023ask}, rewards \cite{leecan}, embodied agents \cite{yang2023octopus}, and adversarial play \cite{xu2023language}. The work focusing on state descriptions \cite{nottingham2023selective, nottingham2023selective2} proposed a method for automatically selecting concise state descriptions for LLM actors in NetHack. They found they could reduce the length of state descriptions by 87\% and improve task success rates by 158\% in NetHack. The research on exploration and intrinsic motivation used background knowledge from LLMs to shape exploration in the Crafter game environment (2D Minecraft) \cite{du2023guiding} and an LLM to construct intrinsic rewards in NetHack \cite{klissarov2023motif}. \citet{du2023guiding} found that their agents had better coverage of common-sense behaviours, while \citet{klissarov2023motif} found that their method significantly outperformed existing approaches. For rewards, \citet{leecan} investigated whether an LLM can postprocess reward signals for RL in MineDojo. They reported that LLMs can suggest a new combination of existing heuristic functions, but did not observe improved learning results. On adversarial play, \citet{xu2023language} proposed a new framework for LLM-based agents with strategic thinking ability to play the language game Werewolf. They found that combining LLMs with the RL policy produced a variety of emergent strategies, achieved a higher win rate, and was robust against human players. The chain-of-thought (CoT) papers found that LLMs prompted with CoT outperformed state-of-the-art RL baselines in Crafter \cite{wu2023spring} and benchmarks such as Overcooked and FourRoom \cite{yan2023ask}. Finally, \citet{yang2023octopus} introduced Octopus, a novel large vision-language model for embodied agents, tested in a custom experimental environment (OctoVerse).

The papers on collaboration and coordination focused on human-AI collaboration \cite{gong2023mindagent, guan2023efficient, brawer2023towards, verma2023preference, liu2023llm} and multi-agent coordination \cite{gong2023mindagent, agashe2023evaluating, chen2023agentverse}. \citet{gong2023mindagent} proposed a new infrastructure, MindAgent, to evaluate planning and coordination capabilities in games for multi-agent systems and human-AI collaboration, and introduced a new test environment, Cuisineworld. \citet{guan2023efficient} proposed using an LLM to develop an action plan to guide both humans and AIs by facilitating a clear understanding of tasks and responsibilities. Their experiments in the Overcooked-AI environment with a human proxy model and real humans found that their method outperformed existing learning-based approaches. \citet{brawer2023towards} proposed a preliminary design for a natural language interface for a task assignment system using an LLM, which they plan to evaluate in the Overcooked environment. \citet{verma2023preference} investigated using LLMs to serve as effective human proxies by capturing human preferences for collaborating with AI agents. They explored the ability of LLMs to model mental states and understand human reasoning processes in Overcooked. \citet{liu2023llm} proposed a hierarchical language agent for human-AI coordination in Overcooked to provide both strong reasoning abilities and real-time execution. They reported that their agent outperformed baseline agents, with stronger cooperation abilities, faster responses, and more consistent language communications. \citet{agashe2023evaluating} introduced the LLM-Coordination (LLM-Co) Framework, designed to enable LLMs to play coordination games, and assessed the effectiveness of LLM-based agents in different coordination scenarios in the Overcooked-AI benchmark. They reported that LLM-based agents can infer a partner's intention and reason actions, coordinate with an unknown partner in complex long-horizon tasks, and outperform RL baselines. \citet{chen2023agentverse} proposed a multi-agent framework that can collaboratively and dynamically adjust its composition as a greater-than-the-sum-of-its-parts system in Minecraft. They found that their framework can effectively deploy multi-agent groups that outperform a single agent and identified the emergence of social behaviours. 

\subsection{Game Development and Play}

Almost a third (25/76, 32.9\%) of the papers related to the theme of Game Development and Play. The papers within this theme related to game and content generation (13/25) \cite{ren2023make, todd2023level, sudhakaran2023prompt, nasir2023practical, kumaran2023end, awiszus2022wor, roberts2022steps, roberts2022surreal, sudhakaran2023mariogpt, taveekitworachai2023chatgpt4pcg, chen2023gamegpt, shams2023towards, colado2023using}, serious games and game-based learning (8/25) \cite{kumaran2023end, colado2023using, saito2023double, tinterri2024ai, junior2023chatgpt, bottega2023jubileo, yue2023combine, amresh2023integrating}, game design (6/25) \cite{lanzi2023chatgpt, huang2023create, saito2023double, tinterri2024ai, junior2023chatgpt, amresh2023integrating}, role-playing games (3/25) \cite{saito2023double, zhu2023fireball, gongora2023skill}, and automated bug detection tools (2/25) \cite{taesiri2023glitchbench, taesiri2022large}. 

The papers on content generation related to game level generation \cite{todd2023level, sudhakaran2023prompt, nasir2023practical, kumaran2023end, awiszus2022wor, roberts2022steps, sudhakaran2023mariogpt, taveekitworachai2023chatgpt4pcg}, game generation \cite{roberts2022surreal, chen2023gamegpt, shams2023towards, colado2023using}, and art generation \cite{ren2023make}. Papers on level generation investigated using LLMs to generate levels for the games Sokoban (2D puzzle game) \cite{todd2023level}, Super Marios Bros (tile-based game levels) \cite{sudhakaran2023prompt, sudhakaran2023mariogpt}, Metavoidal (2D-game rooms for an under-development game) \cite{nasir2023practical}, Future Worlds (strategy game focused on environmental sustainability education) \cite{kumaran2023end}, Minecraft (3D sandbox game) \cite{awiszus2022wor}, VR Pong \cite{roberts2022steps}, and Science Birds (based on Angry Birds) \cite{taveekitworachai2023chatgpt4pcg}. These papers focused on experimenting with controlling LLM level generators \cite{todd2023level}, human-in-the-loop fine-tuning \cite{nasir2023practical}, developing an end-to-end procedural level generation framework \cite{kumaran2023end}, using bootstrapping to generate enough data for training \cite{nasir2023practical}, and establishing a competition environment for participants to compete at creating effective prompts for ChatGPT to generate game levels \cite{taveekitworachai2023chatgpt4pcg}. Researchers found that the performance of LLMs in generating game levels scales dramatically with dataset size, which is consistent with other PCGML (Procedural Content Generation via Machine Learning) techniques, creating the same tension between the cost/benefits of data versus level creation \cite{todd2023level}. They also found that LLMs can be text-prompted for controllable level generation, with varying performance across metrics, addressing a key challenge in current PCG techniques \cite{todd2023level, sudhakaran2023prompt}. Papers on game generation went beyond level generation to look at games holistically or larger game structures/systems. These papers introduced approaches to generate content and mechanics in a VR Pong game during development and play (Codex VR Pong) \cite{roberts2022steps, roberts2022surreal}, a multi-agent collaborative framework (GameGPT) to automate game development \cite{chen2023gamegpt}, and a simulation game system (Infinitia) that uses generative image and language models at play time to reshape the game world (terrain, vegetation, buildings, inhabitants) analogous to the fictional Holodeck \cite{shams2023towards}. Researchers found that LLMs could be used to generate non-deterministic and emergent gameplay \cite{roberts2022steps, roberts2022surreal} and their proposed/developed systems had potential in generating games \cite{roberts2022surreal, chen2023gamegpt, shams2023towards, colado2023using}, but that further work was needed to realise this potential and that human game developers still surpass AI models \cite{roberts2022surreal}. The paper on art generation \cite{ren2023make} aimed to create lifelike 3D avatars from text descriptions using a combination of large language and vision models. The researchers found that their system offered an intuitive approach for users to craft controllable, realistic, and fully-realised 3D characters within 2 minutes, showing potential for integration into a game art pipeline.

The papers on serious games and game-based learning related to using LLMs to support educators in developing games \cite{tinterri2024ai, junior2023chatgpt, yue2023combine, amresh2023integrating}, as well as more specifically to role-playing game design \cite{saito2023double}, board game design \cite{tinterri2024ai, junior2023chatgpt}, and level \cite{kumaran2023end} and game generation \cite{colado2023using} for serious games. A common goal of these papers was to support time-poor teachers in developing educational games for their classrooms in an expedited manner and without the need for expertise in game development \cite{tinterri2024ai, junior2023chatgpt, yue2023combine}. LLMs were used to support educators with brainstorming \cite{junior2023chatgpt}, choosing a game \cite{tinterri2024ai}, personalising the game for constructive alignment and inclusion \cite{tinterri2024ai}, suggesting game themes and mechanisms aligned with curriculum and learning goals \cite{junior2023chatgpt}, providing templates or exemplars of game components \cite{junior2023chatgpt}, tailoring strategies to address specific learning challenges \cite{yue2023combine}, supporting teachers in crafting personalised learning blueprints \cite{yue2023combine}, and offering feedback on game prototypes and identifying areas for improvement \cite{tinterri2024ai}. One paper also explored the creation of games as a learning mechanisms in combination with playing the game, which was found to have “double impact” (first through design, then through play) \cite{saito2023double}. Overall, researchers found that generative AI offers great potential for supporting and enhancing (rather than replacing) designers and educators in the creation of serious games and game-based learning education \cite{colado2023using, tinterri2024ai, junior2023chatgpt}. They also found that generative AI can be an effective design tool for diverse populations (e.g., educators, students, domain experts) \cite{kumaran2023end}.

Apart from the previously discussed papers on design of serious games and game-based learning, the game design papers focused on use of LLMs for game idea generation \cite{lanzi2023chatgpt} and game mechanic design \cite{huang2023create}. The idea generation paper \cite{lanzi2023chatgpt} presented a collaborative design framework that aimed to simulate the typical human design process. As part of this framework, LLMs were used for the recombination and variation of ideas.  The game mechanic paper \cite{huang2023create} used an LLM for an element synthesis game, where players can combine elements to create new content based on physical and chemical properties of the combined elements, conceptual association, combination principles, and other logical reasoning rules. The researchers found that their LLM-based framework was effective in simulating element synthesis in video games with high freedom, high logic, repeatable playability, and more content. However, they noted that the inherent uncertainty in text generated by LLMs can have adverse effects on game mechanics.

The role-playing game (RPG) papers \cite{zhu2023fireball, gongora2023skill, saito2023double} involved using LLMs to design RPGs \cite{saito2023double}, to aid in playing RPGs online \cite{zhu2023fireball}, and act as game masters in RPGs \cite{gongora2023skill}. As previously discussed in relation to serious games, using LLMs to design RPGs was found to be an effective learning tool for children, particularly paired with the children also playing the games they designed \cite{saito2023double}. For aiding people to play RPGs (in this case Dungeons and Dragons, or DnD) online, researchers \cite{zhu2023fireball} collected a dataset with 25,000 unique DnD sessions from real games on Discord that used the Avrae bot, including language, game commands, and underlying game state information. They found that their dataset (FIREBALL) improved natural language generation according to both automated metrics and human judgements of quality. For modelling RPG game masters (GMs), researchers \cite{gongora2023skill} evaluated three LLMs (ChatGPT, Bard, OpenAssistant) as out-of-the-box GMs. Considering the skills needed by a GM (creating and managing a fictional world, tracking the game state, understanding the players’ actions), they found that ChatGPT and Bard provided satisfying game experiences, although they struggled with commonsense reasoning. They found that OpenAssistant was unable to maintain the GM role during most tests.

\subsection{Narrative, Story, and Dialogue}

Most of the remaining papers (17/76, 22.4\%) related to the theme of Narrative, Story, and Dialogue. The papers within this theme related to dialogue or conversation generation (11/17) \cite{nimpattanavong2023fighting, csepregi2021effect, latouche2023generating, muller2023chatter, akoury2023towards, akoury2023framework, volum2022craft, sun2023fictional, paduraruconversational, kumaran2023scenecraft, ashby2023personalized}, story generation and interactive story (6/17) \cite{taveekitworachai2023waiting, gursesli2023chronicles, taveekitworachai2023journey, yong2023playing, sun2023language, akoury2023towards, kumaran2023scenecraft}, and quest generation (2/17) \cite{ashby2023personalized, vartinen2022generating}. Two papers related to both dialogue and story generation \cite{akoury2023towards, kumaran2023scenecraft} and one paper to both dialogue and quest generation \cite{ashby2023personalized}. Two papers considered both NPC dialogue and NPC actions/tasks \cite{volum2022craft, paduraruconversational}.

The papers on dialogue generation mostly focused on non-player characters (NPCs) in games. However, there was one case where the dialogue was for an online streamer commentating matches in a fighting game (DareFightingICE) \cite{nimpattanavong2023fighting} and another where NPCs from a game were interacting with players in an online Discord community external to the game \cite{sun2023fictional}. Papers on dialogue generation focused on different dimensions of dialogue and NPCs, including context-awareness and in-context dialogue \cite{csepregi2021effect, muller2023chatter, ashby2023personalized}, style and personality \cite{latouche2023generating, muller2023chatter}, and story-focused dialogue \cite{akoury2023towards, kumaran2023scenecraft}. Most papers performed an evaluation of their systems with human players or game developers. The reported results were mixed, depending on the focus of the papers and the criteria for success. Some researchers reported on the difficulty of using LLMs to generate NPC dialogue \cite{akoury2023towards} and evaluator preferences for game designer generated text compared to LLM-generated text \cite{akoury2023framework}. Others reported that the LLM-generated text enhanced the player experience generally \cite{ashby2023personalized, paduraruconversational, sun2023fictional} or within specific conditions (e.g., with context-aware NPCs \cite{csepregi2021effect}). Some papers reported on the success of the approach via other performance metrics \cite{latouche2023generating, muller2023chatter, volum2022craft, kumaran2023scenecraft}.

The papers on story and quest generation investigated different aspects of game story endings \cite{taveekitworachai2023waiting, gursesli2023chronicles, taveekitworachai2023journey}, the use of LLMs for generating interactive story games \cite{yong2023playing, sun2023language}, and quest description generation \cite{ashby2023personalized, latouche2023generating}. Papers focusing on story endings explored the valence (positive/neutral/negative) \cite{taveekitworachai2023waiting} and bias of the generated endings \cite{taveekitworachai2023journey} and the impact of generating stories with/without a prompted ending (along with inspiration keywords) \cite{gursesli2023chronicles}. The interactive story generation papers studied players interacting with the commercial game AI Dungeon \cite{yong2023playing} and a custom game "1001 Nights" \cite{sun2023language}. Research related to story endings found that generated stories were biased towards positive endings \cite{taveekitworachai2023journey}, that models classified stories into uninstructed endings \cite{taveekitworachai2023waiting}, and that more detail in prompts led to higher coherence but less inspiration \cite{gursesli2023chronicles}. The interactive story researchers observed that player's mental models shifted over the play session from assuming a linear-branching narrative to one with open possibilities, impacting their motivations for repeat engagement \cite{yong2023playing}. One paper proposed the term "AI-Native games" for games where generative AI is fundamental to the game's mechanics \cite{sun2023language}, also noting that inconsistency, incoherence, and  AI-transparency are key challenges in these games. One paper focused on quest generation found that LLM-generated quests approached hand-crafted quests in terms of fluency, coherence, novelty, and creativity according to human evaluation \cite{ashby2023personalized}. Conversely, the other quest generation paper found that only 1 in 5 generated quests were deemed acceptable by a human critic, with a large variation in quality of the generated quests \cite{latouche2023generating}.

\subsection{Game Research and Reviews}

The remaining papers (7/76, 9.2\%) related to LLMs for use in analysing or generating data about games, including research data and game reviews. The papers in this theme used LLMs to analyse game reviews \cite{lankes2023game, viggiato2023leveraging} and interviews with video game players \cite{de2023performing} and to generate synthetic data for hate speech in games \cite{svetasheva2024harnessing}, player responses to interview questions \cite{hamalainen2023evaluating, hamalainen2022neural}, and utterances about video games \cite{ramirez2023controlling}. Researchers found the LLMs to be effective for their research purposes in analysis and generation, including dealing with the complex and highly abbreviated lexicons related to games \cite{svetasheva2024harnessing} and offering improvements in performance \cite{viggiato2023leveraging} and cost \cite{hamalainen2023evaluating, hamalainen2022neural}. However, researchers also noted some drawbacks, including variation in quality \cite{hamalainen2022neural}, lack of depth of content and flexibility \cite{lankes2023game}, and potential for crowd-sourced participant data to become unreliable \cite{hamalainen2023evaluating}. 

\subsection{Recommendation}

During our search process, we found an additional 9 papers \cite{yue2023llamarec, yang2023large, zhang2023recommendation, bao2023bi, lei2023recexplainer, zheng2023adapting, shu2023rah, sun2023large, charity2023preliminary} that used game data sets as part of broader LLM for recommendation (LLM4Rec) research. These papers did not fit the focus of our search, or of this preliminary scoping review. However, we felt it was important to capture them in this paper, given their proximity to the topic and their likely relevance to the games research community in future. Each of these papers made use of a games dataset (Amazon 7/9, Steam 3/9), alongside other datasets (e.g., movies, books). The LLM4Rec papers focused on performance of ranking-based recommendation \cite{yue2023llamarec, bao2023bi}, sequential recommendation \cite{yang2023large}, instruction following and tuning \cite{zhang2023recommendation}, interpretation and explanation \cite{lei2023recexplainer}, collaborative semantics \cite{zheng2023adapting}, user alignment \cite{shu2023rah}, intent-aware session recommendation \cite{sun2023large}, and game features \cite{charity2023preliminary}. The LLM4Rec papers most frequently utilised LLaMA (4/9) and GPT (4/9), whereas the games-focused papers overwhelmingly made use of GPT.

\section{Conclusions}
We reviewed 76 papers on LLMs and video games, published between 2022 to 2024, in order to provide a first snapshot of the state of research on this topic and to lay the foundations for the rapidly moving ongoing and future work. We designed our review to be narrowly focused on the application of LLMs to video games. As noted earlier, we included papers that reported original research related to video games and LLMs. We excluded work that had not yet been applied to games (but that could be applied to games in future). We also did not explicitly search for areas related to games (e.g., gamification) or other types of generative models (e.g., image generation, multi-modal generation). We found that research clustered around topics related to game AI (agents, RL, collaboration), game development (content/game generation, serious games and game-based learning, game design, RPGs, bug detection), narrative (conversation/story/quest generation), and game research and reviews (analysing/generating game data). Much of the work reviewed constituted initial attempts at applying LLMs to different aspects of games. Researchers generally reported positive results or indicated promising future work related to their application of LLMs to games. Others reported success in combining LLMs with other approaches (e.g., RL) to outperform baselines. Some of the commonly reported highlights included human interpretability, social behaviours and experiences, foundational skills (e.g., can play games), and empowerment of non-developers to create games. Negatives included lack of logical reasoning ability and unpredictability (e.g., in the context of generating content in live games). Given the explosion of interest around LLMs, research will continue to move quickly on this topic. Our review aimed to provide a preliminary snapshot of the state of the work on LLMs and games, to help researchers to understand the key directions and findings so far, and to lay the foundation for future work. We expect that capturing the state of the field and keeping pace with new research will be an ongoing challenge, but it is exciting to see where LLMs will take the field of games research and development.

\bibliographystyle{ACM-Reference-Format}
\bibliography{arxiv}

\end{document}